\documentclass[a4paper,10pt]{article}

\usepackage{graphicx}

\begin{document}


\begin{center}

\Huge{Deflection of Molecules by a Homogeneous Electric Field: A New Effect}

\end{center}

\small

%


%

\renewcommand{\thefootnote}{\fnsymbol{footnote}}

{\bf M. Dorado{\footnote[1]{To whom correspondence should be addressed. 

E-mail:  mdorado@cayacea.com}}, M. Morales Furi\'o,

J.L. P\'erez Fern\'andez and J.L. S\'anchez G\'omez{\footnote[2]{Permanent address: 

Dept. F\'{\i}sica Te\'orica U.A.M., Cantoblanco, 28049-Madrid}}} \\

Centro de Investigaci\'on de T\'ecnicas Aeroespaciales, S. A.\\

(CENITA, S.A.) \\

C/ Miguel Yuste, 12, 28037 Madrid, Spain

\vspace*{0.5cm}

\noindent

\textbf{Abstract} \\

In this work we put forward a theoretical explanation of a peculiar effect
found very recently by A. Gonz\'alez Ure\~na {\it et al.} \cite{3}. They have 
observed the deflection of a beam of molecules posessing a permanet 
electric dipole moment by a {\it homogeneous} electric field when a resonant
oscillating field is superposed transverse to the static one.

\normalsize

\hspace*{1cm}

\section*{\normalsize INTRODUCTION}

The molecular beam technique has significantly contributed 
to the development of the Atomic and Molecular Physics. Well 
known examples of this intellectual and scientific development
 are the molecular beam magnetic or electric resonance 
spectroscopy. There are excellent reviews \cite{1,2} on the 
subject emphasizing either spectroscopic or scattering 
applications. The basic principle exploited in these techniques 
is the magnetic or electric focussing and/or depletion due to 
the interaction between the permanent magnetic or electric 
moment of a molecule and a non-homogeneous magnetic or electric 
field, respectively.

The aim of the present work is to put forward an explanation 
to the effect reported in ref. \cite{3}

In that experiment a N$_{2}$O supersonic beam that interacts 
with both a DC Stark field and a resonant radiofrequency 
radiation. During the flight of the molecular beam pulse 
from the source to the detector, the molecules pass through 
a homogeneous static electric field. Changes in the beam 
intensity were observed when a resonant oscillating electric 
field is superposed transverse to the static field. Thus, a 
new phenomenon was observed which consists of N$_{2}$O beam 
depletion upon its interaction with both a homogeneous electric 
field and a resonant radiofrequency, see ref. \cite{3} for details.

\section*{\normalsize THEORETICAL DISCUSSION}

When a neutral particle with an electric dipolar moment, 
$\mathbf{\mu}$, goes through an electric field, $\mathbf{E}$,
upon the particle is acting a force

\begin{equation}
   \mathbf{F} = \mathbf{\nabla}(\mathbf{\mu}\cdot \mathbf{E})
\end{equation}

Hence, if the electric field is homogeneous the corresponding 
force vanishes and the particle's trajectory will not be affected 
by the electric field.

For understanding the experimental results\cite{3} two types 
of interaction are of relevance. Namely, the interaction of the
permanent dipole moment with the homogeneous electric field  
(Stark interaction) and the new interaction between the particle 
and both the static and the resonant oscillating field (depletion 
interaction). Let us consider them briefly:

\vspace*{0.5cm}

\noindent

(a) \emph{Stark interaction}

The quantum levels of a molecule that enters a homogeneous 
electric field are split by the Stark effect. Then, a direction 
`that of the field' is singled out.

The rotational energy levels of a molecule (rigid rotor) are 
perturbed and given by\cite{4}:
\begin{equation}
   W_{JM} = hB\left[J(J+1) + \left(\frac{\mu}{hB}\right)^2\frac{J(J+1)-3M^2}{2J(J+1)(2J-1)(2J+3)}E^2\right]
\end{equation}

Where $J$ is the rotational quantum number, $E$ is the electric
field, $M$ is the component of the angular momentum
along the quantization axis, $<\mu>$ the 
permanent electric dipole moment of the molecule, and $B$ its
rotational constant. This equation is obtained by the  
second-order perturbation theory\cite{4}. (The odd terms in 
the series are identically zero and the fourth term is neglected 
because it is quite small at the conditions of the present 
experiment\cite{5}). As mentioned earlier the transition 
considered in the experiment is the (1,$\pm$1) $\rightarrow$ (1,0). 
For this transition  equation (2) yields an energy splitting, $\Delta W$:

\begin{equation}
   \Delta W = W(1,0) - W(1,\pm 1) = \frac{3}{20}hB\lambda^2, \; \; \mbox{with} \; \; \lambda=\frac{\mu E}{hB}
\end{equation}

The oscillator frequency $\omega$ for the transition is 
obtained applying the usual Bohr relation 

\begin{equation}
   \omega=\Delta W/\hbar
\end{equation}

For example, for the N$_{2}$O molecule, with $<\mu>$ = 0.161 D 
and I=6.68 10$^{-46}$ Kg m$^{2}$ \cite{6,7,8}, the frequency of 
the transition (1,$\pm$1) $\rightarrow$ (1,0)is equal to 286.9 kHz 
when the static field has an intensity of 191.3\ kV·m$^{-1}$.

\vspace*{0.5cm}

\noindent

(b) \emph{Depletion interaction}

In a previous study on the dynamics of a moving system, which 
also has an internal angular momentum\cite{9}, the main conclusion 
of that work was that the trajectory of a moving particle, 
possessing internal angular momentum, can be modified when a 
torque is applied with a perpendicular component to the rotational
angular momentum vector.  Specifically the following behaviour was 
predicted\cite{5} for a `particle' with angular momentum $\mathbf{J}$
and velocity $v$ to which a torque $\Gamma$ is applied the
system would behave as if a central force given by equation (5) 
applied. See Appendix for the main points of the derivation and 

ref. \cite{9} for details. 

\begin{equation}
   \mathbf{F} = m \mathbf{v} \times \mathbf{\Omega}
\end{equation}
where $\mathbf{\Omega}=\frac{\mathbf{\Gamma}}{<\mathbf{J}>}$ and 

where $\mathbf{\Omega}$ direction, is that of external field.

To test this force, an experiment for a microscopic system was 
suggested\cite{9}, based on the interaction of a homogeneous magnetic 
field and  electromagnetic radiation with a particle that has spin and 
magnetic moment. It was then predicted that the deviation from the 
original trajectory should take place when the radiation is resonant 
with a quantum transition between the energy levels of the particle.

Notice that according to the current wisdom (i.e. standard  mechanics 
and electromagnetic ) no deviation of the particle should occur as it 
is moving in a homogenous magnetic field.

The present work deals with an experiment carried out with a N$_{2}$O 
supersonic beam that interacts with both a DC Stark field and a resonant 
radiofrequency radiation. From a physical point of view the (present) 
electric resonant beam experiment is equivalent to the one suggested 
in Ref \cite{9}. Indeed, a Stark field replaces the Zeeman field and 
the particle magnetic moment is replaced by the permanent dipole moment 
of the linear N$_{2}$O molecule, which for the present investigation is 
considered as a rigid rotor. Again according to standard mechanics and 
electromagnetism the trajectory of a molecule in the beam should be not 
affected by the electric field since this is homogeneous. So that, 
any eventual beam depletion would in principle indicate the presence 
of the new force mentioned above (unless some other `conventional' 
interpretation, of which we are not aware, could be found).

As said above, we deal with the rotational quantum states $J$ = 1 $M_{J}$= 0,
$\pm$ 1. The Stark interaction ensures molecular polarization  
and therefore quantized orientations of the $J$ vector with respect to
$\mathbf{E}$ while the molecule keeps rotating at each of these polarized
($M_{J}$) states.

We will deal with the motion of the molecular beam under the applied 
fields, $\mathbf{E}$, homogeneous and constant, and $\mathbf{E_1}$,
which is an oscillating field. To do it, we are going to use a 
semiclassical approach wherein it is assumed that such motion can be 
reasonably well described by studying how the mean values (for the 
relevant quantum states) of the dipole moment operator evolve in 
presence of the external fields and then applying Eq. (5) with the 
torque computed by using those mean values of the dipole operator.

Now when no external field is present the mean value of the dipole operator, 
$\mathbf{\hat\mu}$
, is zero for any rotational state, that is 
$<\psi_{JM}|\mathbf{\hat\mu}|\psi_{JM}>=0$
, simply due to symmetry (parity) reasons. But if the molecule is 
interacting with a homogeneous electric field, $\mathbf{E}$=(0,0,$E$), there
is a interaction hamiltonian

\begin{equation}
   \hat H_{int} = - \mathbf{\hat\mu} \cdot \mathbf{E} = -<\mu>E \cos\theta
\end{equation}
where $\theta$ is the polar angle, and then $J$ is no longer a
`good' quantum number since $[\hat H_{int}, \mathbf{\hat J}] \not=0$.

Restricting our discussion to the relevant, $J$=1, non perturbed
states, and in first order perturbation theory, the corresponding 
quantum states with $\mathbf{E}$ present are

\begin{eqnarray}
  |\Psi_{1M}> & = & |\psi_{1M}>+\frac{<\psi_{00}|H_{int}|\psi_{1M}>}{\varepsilon_1 - \varepsilon_0}|\psi_{00}>\delta_{M0} \nonumber \\ 
& + & \frac{<\psi_{2M}|H_{int}|\psi_{1M}>}{\varepsilon_1 - \varepsilon_2}|\psi_{2M}> \; \; (M=0,\pm1)
\end{eqnarray}

Where $|\psi_{JM}>$ are the `non perturbed' (rotational) states and 
$\varepsilon_J=hBJ(J+1)$ are the unperturbed (no field) energies. 
(Note that, in spite of the notation, $|\Psi_{1M}>$ 
 is not an eigenstate of $\mathbf{\hat J^2}$; yet it is of $\mathbf{\hat J_Z}$, with 
eigenvalue $M$.)

Now the mean value of $\mathbf{\hat \mu}$ in a state $|\Psi_{1M}>$
is no longer zero. In fact, one has

\begin{equation}
   <\Psi_{1M}|\hat \mu_X|\Psi_{1M}> 
   = <\Psi_{1M}|\hat \mu_Y|\Psi_{1M}> = 0
\end{equation}
but

\begin{equation}
   <\Psi_{1M}|\hat \mu_Z|\Psi_{1M}> = - \frac{W_{1M}-\varepsilon_1}{E}\not=0
\end{equation}
where W$_{1M}$ is the energy corresponding to the state $|\Psi_{1M}>$ 

 (Eq. (2))

Therefore, we see that, in presence of the constant field $\mathbf{E}$,
the mean value of the dipole operator in any quantum
state (that is, for all values of M=0,$\pm$1) is a vector directed along 
the field (Z) direction.

In this situation, the new effect does not operate because the 
corresponding torque (computed as stated before) is zero, and, 
consequently, no depletion of the molecular beam takes place.

However, the presence of an oscillating electric field, 
$\mathbf{E_1}=(E_1 \cos \omega t, E_1 \sin \omega t,0)$, perpendicular to 
$\mathbf{E}$, as shown in Figure 1, where this situation 
is been displayed just before the interaction takes place, gives 
rise to a torque, $\mathbf{\Gamma_1}=<\mathbf{\mu}>\times \mathbf{E_1}$ (where 
$<\mathbf{\mu}>$ is the mean value of $<\mathbf{\hat \mu}>$
 in the corresponding quantum state), that makes the molecule´s 
dipole (its mean value, of course) to deviate from its previous 
direction (parallel to $\mathbf{E}$) due to the nutation performed 
by the molecule´s angular momentum, the nutation velocity being 

(recall eq. (5))

\begin{equation}
   \Omega_1 = \frac{\Gamma_1}{<J>}
\end{equation}
where $\mathbf{\Omega_1}$ direction, is that of $\mathbf{E_1}$.

Then, there is a torque induced by $\mathbf{E}$, $\mathbf{\Gamma_2}=<\mathbf{\mu}>\times \mathbf{E}$ 
(because $\mathbf{E}$ and $<\mathbf{\mu}>$ are no longer parallel ) and, 
besides the said nutation, the angular momentum, $<\mathbf{J}>$, 
starts  precessing  about $\mathbf{E}$ with a velocity

\begin{equation}
   \Omega_2 = \frac{\Gamma_2}{<J>}
\end{equation}
where $\mathbf{\Omega_2}$ direction, is that of $\mathbf{E}$.

If the frequency of the precession induced by the homogeneous 
electric field is resonant with  the frequency of the oscillating 
field, then the interaction keeps up coherence, that is, the plane 
defined by the homogeneous field $\mathbf{E}$ and the dipole moment
$<\mathbf{\mu}>$ (to be called $\alpha$ hereafter) precesses about $\mathbf{E}$ as
depicted in Figure 2. In this Figure, the system evolution is shown 
after a time t, together with the trajectory S followed by the molecule 
under resonant conditions. Notice how the oscillating field, 
$\mathbf{E_1}$ (which is emulated experimentally by an electromagnetic
radiation),  remains perpendicular  to $\alpha$ during the interaction 
time. Indeed, the particle deflection from the original X direction is 
clearly noticed.  On the other hand, if there is no resonance, the 
$\alpha$ plane precession velocity, $\mathbf{\Omega_{2}}$, and that of the 
oscillating field, $\omega$, are different, and  then the interaction 
is not coherent and the effect (averaged in time) turns out to be 
null. (Notice that, from (10), one has 
$\Omega_1=\frac{<\mu> E_1 \cos(\Omega_2-\omega) t}{<J>}$ ).

Then (see Appendix and for a complete classical treatment, ref. \cite{5}) 
there will be  two forces (of analogous type) acting  at the CM of 
the molecule that will modify its trajectory (recall eq. (5))

\begin{eqnarray}
   \mathbf{F_1} & = & m \mathbf{v} \times \mathbf{\Omega_1} \nonumber \\
          \\
   \mathbf{F_2} & = & m \mathbf{v} \times \mathbf{\Omega_2} \nonumber 
\end{eqnarray}
where  $\mathbf{v}$ is the velocity of the centre of mass of the molecule. 

When the interaction begins, and because of  the oscilating field, 
molecules in resonance start to follow a circular trajectory in  
plane $\alpha$ , of radius $R_1=\frac{v}{\Omega_1}$.

Simultaneously, $\mathbf{F_{2}}$, causes the precession of 
plane $\alpha$, that contains the circular trajectory about 
$\mathbf{E}$. This interaction can be visualized by regarding the
said molecules as describing a circle in $\alpha$  while this plane 
is precessing about $\mathbf{E}$ with velocity $\mathbf{\Omega_{2}}$. This
effect remains as long as such molecules are acted by both fields, 
and, consequently, they will not reach the detector.

To summarize: If the oscillating electric field $\mathbf{E_{1}}$ is not 
resonating, that is its frequency is not that corresponding to 
the transition between the $M_{J}$ = 0 and $M_{J}$ = $\pm$ 1 
levels, then   the new   effect is  absent and, consequently, no 
depletion of the molecular beam takes place  . However, things 
are different when the oscillating field has a frequency which
 corresponds to the transition between the mentioned levels, for, 
in such a case, and as predicted in \cite{9} an effect appears 
which deflects the trajectory of molecules in the $J$=1 state, as 
observed in the present experiment. The depletion depends on both 
the intensity of the homogeneous electric field and the resonant 
field. Nevertheless, the important point is that the oscillation 
between the two levels is the necessary ingredient for the new 
interaction totake place, not the transition between levels itself.

\section*{\normalsize CONCLUSION}

The effect reported in ref. \cite{3} can be understood by a 
semiclassical model, presented here, based upon the work 
presented in ref. \cite{9}, which concerns only the classical 
domain. Up to our knowledge, there is no conventional explanation 
for this effect.

\newpage

\section*{\normalsize Figures}

\begin{center}

\includegraphics[width=15cm, height=15cm]{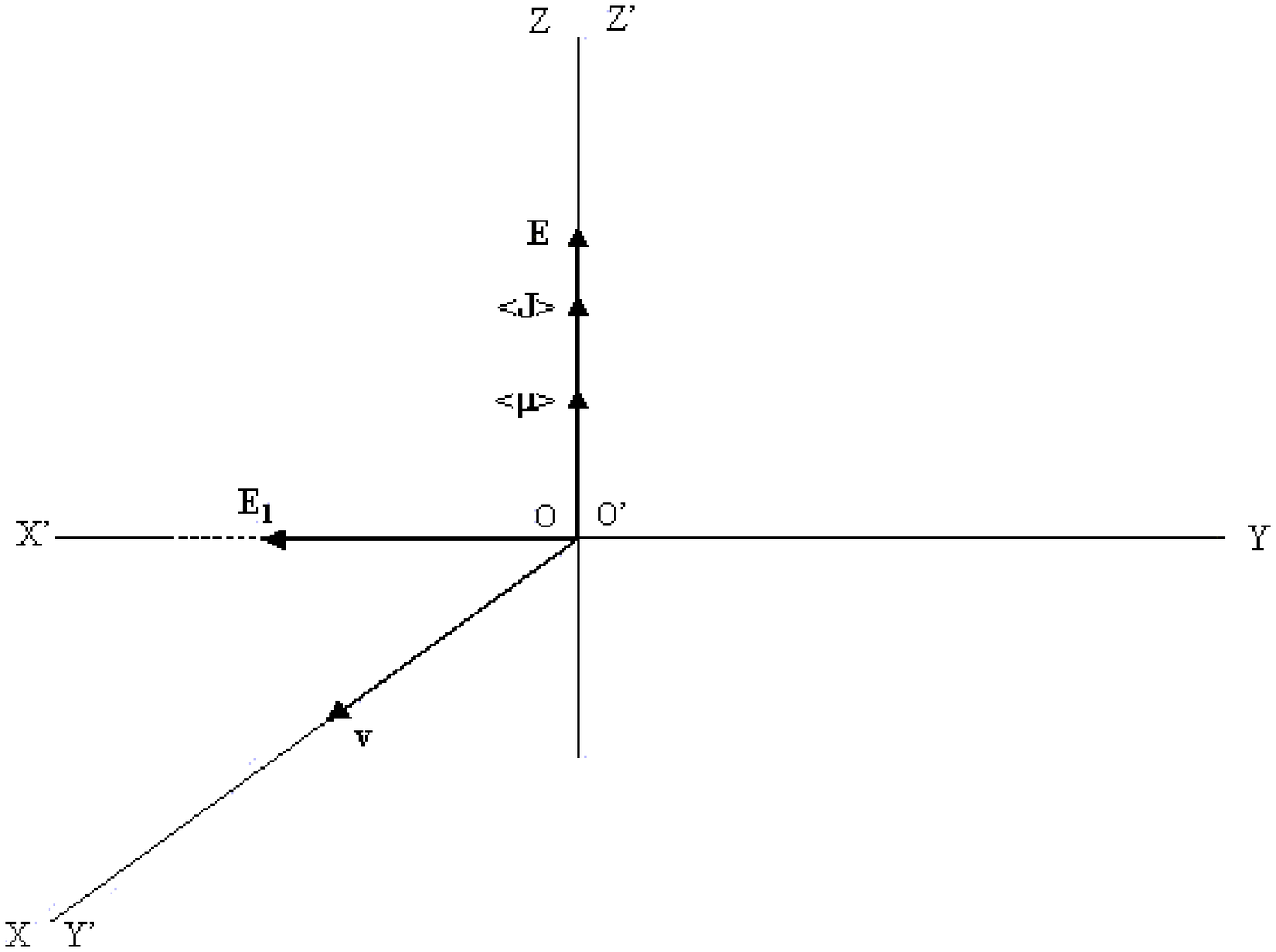}

\end{center}

{\bf Figure 1:} The frames X, Y, Z (laboratory), and X'Y'Z' (attached 
to the molecule) are shown. O is the origin of the laboratory 
system; O', that of system axis linked to the particle and also 
the location of the centre of mass of the particle. The mean 
dipole moment $<\mathbf{\mu}>$ and angular momentum $<\mathbf{J}>$ 
are parallel to $\mathbf{E}$ (Z-axis). The situation is represented
when, due to the interaction with $\mathbf{E_1}$ the depletion
of the molecule starts.

\begin{center}

\includegraphics[width=15cm, height=15cm]{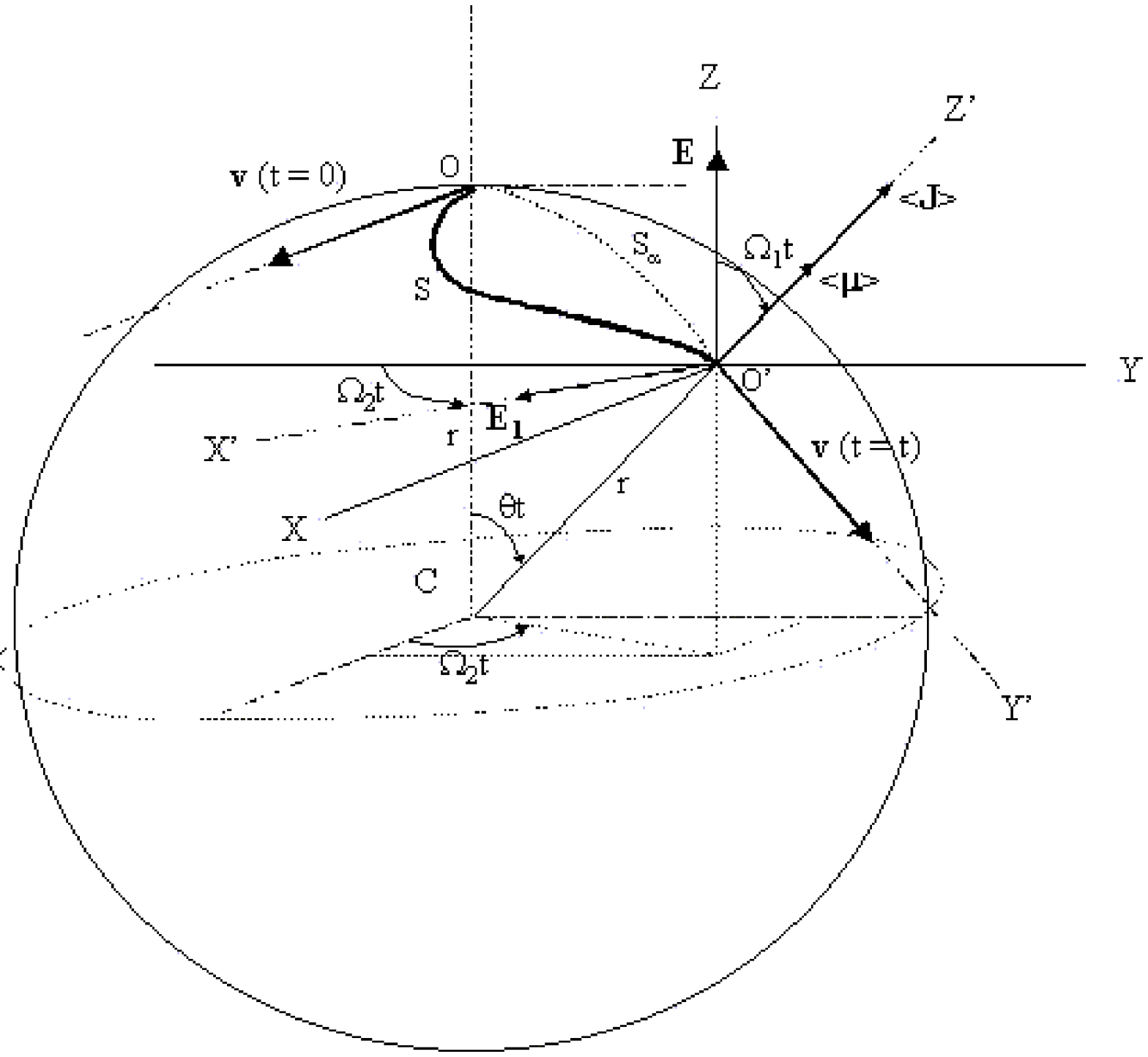}

\end{center}

{\bf Figure 2}: Detailed view of the particle's trajectory 
from t = 0 until t = t when the resonant conditions are fulfilled. 
The molecule moves from O along the spherical surface with radius 
r centred at C. S, the particle's trajectory results from the 
combinations of (a) a circular path, S$_{\alpha}$, with radius 
r and centred at C along the Y'Z' plane, called $\alpha$ hereafter, 
wich contains $\mathbf{E}$, $<\mathbf{\mu}>$, $<\mathbf{J}>$ and $\mathbf{v}$, and
(b) the rotation of $\alpha$ around the OC axis with (angular) 
velocity  $\mathbf{\Omega_{2}}$. The XYZ frame has been displaced to O' 
for a better illustration of the rotational movements. The X'Y'Z' 
plane is tangent to the sphere at O'; consequently the Z' axis is 
perpendicular to the surface of the sphere at O' and contains C. 
As indicated, the particle's trajectory is always located on the 
sphere surface. Since the position of O' is defined by $\theta$ and r,
$\dot{\theta}$ defines the O' angular velocity with respect to 
the laboratory axis. In addition $\mathbf{\Omega_{1}}$ is the X'Y' 
(nutation) velocity with respect to the laboratory axis. Notice how
$\dot{\theta}$ and $\mathbf{\Omega_{1}}$, always coincide as described 
in equation (A11). See text and appendix for further details.

\begin{appendix}

\section*{\normalsize APPENDIX}

In this Appendix we present a brief treatment of the classical 
motion of a rigid body- with internal angular momentum -- that is 
moving with respect to a given (inertial) frame and is subjected to 
the action of a torque which is perpendicular to the angular momentum 
vector. This situation is clearly related to the classical motion of 
molecules as described in the text.

The corresponding equations of motion will be obtained by the 
Hamiltonian formulation. It should be noted that the Hamiltonian 
formulation is developed for holonomous systems and the forces 
derived from a potential that depends on the position or from 
generalized potentials. A torque is applied to the present system. 
The result of the forces is zero on the center of mass and, therefore, 
it is meaningless to refer to potential energy.

The Hamiltonian describing the centre of mass motion of the `body' 
is, therefore, that of a free particle but for the fact that the 
motion is constrained by the presence of the torque. Such constrain 
is expressed by the equation

$$
  \left(\frac{d\mathbf{J}}{dt}\right)_{XYZ}=\left(\frac{d\mathbf{J}}{dt}\right)_{XYZ} + \mathbf{\Omega} \times \mathbf{J}
\eqno(\mbox{A}1)
 $$
in which $\mathbf{\Omega}$ is the rotation velocity of the frame 
linked to the particle (X',Y',Z') about the frame of inertial 
reference axes (X,Y,Z)  (see \cite{10})

We use polar coordinates in the plane of motion
$$
   v_r=\dot{r} \; \mbox{;} \; v_{\theta}=r\dot{\theta} \; \mbox{;} \; v_Z=\dot{Z}
\eqno(\mbox{A}2)
$$
and the corresponding momenta
$$
  P_r=m\dot{r} \; \mbox{;} \; P_{\theta}=m r^2\dot{\theta} \; \mbox{;} \; P_Z=m\dot{Z}
\eqno(\mbox{A}3)
$$

The Hamiltonian is then
$$
   H = \frac{P_r^2}{2m}+\frac{P_{\theta}^2}{2mr^2}+\frac{P_Z^2}{2m}
\eqno(\mbox{A}4)
$$

The Hamilton equations are

$$
  \dot{r}=\frac{\partial H}{\partial P_r} = \frac{P_r}{m} \; \mbox{;} \; \dot{\theta}=\frac{\partial H}{\partial P_{\theta}} = \frac{P_{\theta}}{mr^2}\; \mbox{;} \; \dot{Z}=\frac{\partial H}{\partial P_Z} = \frac{P_Z}{m} 
 \eqno(\mbox{A}5)
$$
$$
  -\dot{P}_r=\frac{\partial H}{\partial P_r} = -\frac{P_{\theta}^2}{mr^3} \; \mbox{;} \; -\dot{P}_{\theta}=\frac{\partial H}{\partial \theta} = 0 \; \mbox{;} \; -\dot{P}_Z=\frac{\partial H}{\partial Z} = 0
 \eqno(\mbox{A}6)
$$

>From these equations (A6 it is shown that the angular momentum  orbital 
$P_{\theta}$ is conserved.
$$
   P_{\theta} = P_Z = \mbox{const.}
\eqno(\mbox{A}7)
$$
The first equation in (A6) gives the radial equation of motion 
from (A3) it results that
$$
   \dot{P}_r = \frac{m^2 r^4 \dot{\theta}}{mr^3} = m r \dot{\theta}^2
\eqno(\mbox{A}8)
$$

The term $-\frac{\partial V}{\partial r}$ normally appears in 
this radial equation of motion and represents the force derived 
from a potential. In the present case, this term does not exist.

Now we incorporate the constrain, Eq. (A1), taking into account that
$\left(\frac{d\mathbf{J}}{dt}\right)_{X'Y'Z'}=0$, because by assumption 
the torque is external (with respect to the body), and 
$\mathbf{\Gamma}=\left(\frac{d\mathbf{J}}{dt}\right)_{XYZ}$
Then we have

$$\left(\frac{d\mathbf{J}}{dt}\right)_{XYZ} = \mathbf{\Omega} \times \mathbf{J}
\eqno(\mbox{A}9)$$
>From this,
$$\Omega = \frac{\Gamma}{J}
\eqno(\mbox{A}10)$$
\textbf{$\Omega$} direction is that of the electric field 
that produces the torque.

In polar coordinates, the variable defining the rotation about 
the system of axes is $\dot{\theta}$ and it is concluded that

$$ \dot{\theta} = \Omega = \frac{\Gamma}{J}
\eqno(\mbox{A}11)$$

substituting in the radial equation of motion, one obtains:

$$ m\ddot{r}=m r \dot{\theta}^2 = m r \Omega^2
\eqno(\mbox{A}12)$$

>From (A3) it results that the initial moments equals,

$$ P_{\theta} = m r^2 \dot{\theta} = mrv
\eqno(\mbox{A}13)$$
as v is constant, it is concluded that r is also constant. 

Finally one gets 
$$ r = \frac{v}{\Omega} \; \mbox{and} \; m\ddot{r}= mv \Omega
\eqno(\mbox{A}14)$$

A geometric treatment lets to conclude that:
$$\mathbf{F} = m\mathbf{v} \times \mathbf{\Omega}
\eqno(\mbox{A}15)$$

\end{appendix}


\begin{thebibliography}{10}
\bibitem{5} As it is well known the N$_{2}$O linear molecule, 
when excited in its bending mode, acquires an angular momentum 
along the molecular beam axis, which leads to a linear Stark effect. 
Nevertheless at the conditions of the present experiment these 
transition lie within the Megahertz range, i.e. far from the 
radiofrequency values used here.
\end{thebibliography}
\end{document}